\documentclass[]{raa}

\usepackage{graphicx,times}  
\usepackage{amsmath}
\usepackage{amsfonts}
\usepackage{amssymb}
\usepackage{natbib}
\usepackage{comment}
\bibpunct{(}{)}{;}{a}{}{,}
\usepackage[pagebackref=true]{hyperref}

\begin{document}

\newcommand{\obj}{J0026}

  \title{The Optical Study of the Eclipsing Polar SDSS J002637.06+242915.6}

   \volnopage{Vol.0 (202x) No.0, 000--000}      
   \setcounter{page}{1}          

   \author{V. Yu. Kochkina 
      \inst{1,*}\footnotetext{$*$Corresponding Author.}
   \and A. I. Kolbin
      \inst{1}
   \and T. A. Fatkhullin
      \inst{1}
   \and A. S. Vinokurov
      \inst{1}
   \and N. V. Borisov
      \inst{1}
   }

   \institute{Special Astrophysical Observatory of the Russian Academy of Sciences, Nizhnii Arkhyz, Karachai-Cherkessian Republic, Russia; {\it nikainspace@gmail.com}\\
\vs\no
   {\small Received 202x month day; accepted 202x month day}}

\abstract{ We have analyzed phase-resolved photometric and spectroscopic observations of the eclipsing polar SDSS J002637.06+242915.6. The light curve has a M-shaped bright phase that was reproduced using a simple model of an accreting magnetic white dwarf. The hydrogen emission lines exhibit a narrow component formed on the irradiated hemispere of the donor. The Doppler tomography revealed differences in the positions of emission regions of hydrogen and HeII~$\lambda$4686 lines. The spectra exhibit a Zeeman absorption triplet of the H$\alpha$ line, formed in the cold halo around the accretion spot at a magnetic field strength of $B = 15.1 \pm 1.3$ MG. The spectra of the bright phase have a red cyclotron continuum, whose orbital variability has been interpreted within a simple model of an accretion spot. The modeling of the cyclotron continuum constrains the white dwarf's magnetic field to $B_{cyc} \lesssim 45$~MG. The analysis of the eclipse light curve and the radial velocities of the irradiated hemisphere yielded estimates for the orbital inclination $77.2^\circ \le i \le 80.6^\circ$, the mass ratio $0.23 \le q \le 0.43$, and the white dwarf mass $0.72 \ge M_1/M_\odot \ge 0.42$.
}

   \authorrunning{Kochkina et al.}            
   \titlerunning{Eclipsing Polar SDSS J002637.06+242915.6}  

   \maketitle

\section{Introduction}

Cataclysmic variables are close binary systems consisting of an accreting white dwarf and a low-mass star filling its Roche lobe \citep{Warner95}. Among them, a subclass of polars (or AM~Her-type stars) is distinguished by the high magnetic field strength of the white dwarf ($B\sim 10-100$~MG). The magnetic field in polars prevents the formation of an accretion disk. The accretion stream flows ballistically from the inner Lagrangian point L$_1$ before magnetic threading in polars, subsequently being directed by magnetic field lines toward one or both magnetic poles. The impact of infalling gas onto the surface of the accretor results in the formation of hot ($T \sim 10-50$~keV) accretion spots \citep{Cropper90}. They are bright sources of X-ray emission ($L_X \sim 10^{31} - 10^{32}$~erg s$^{-1}$, \citealt{Mukai17}), as well as sources of polarized cyclotron radiation in the optical and infrared ranges. The strong magnetic field of polars makes them synchronous systems, where the spin period of the white dwarf is strictly equal to the orbital period.

The study of polars is important for several reasons. First, polars are valuable objects for understanding the influence of magnetic fields on the evolution of close binary systems. For instance, there are observational indications of differences in the evolution of polars and non-magnetic cataclysmic variables \citep{Belloni20, Shreiber24}. Second, research on polars is necessary for understanding the origin and structure of the magnetic fields of accretors in cataclysmic variables \citep{Ferrario15, Briggs18}. Third, polars are the natural laboratories for studying the interaction of supersonic plasma with the magnetospheres of accretors \citep{Hameury86, Li99}.
Currently, a fairly large number of AM~Her-type systems are known ($\approx250$, \citealt{Schwope25}), but among them, eclipsing systems are particularly valuable. The presence of eclipses makes it possible to determine the parameters of the binary system, which are very difficult to ascertain in polars due to the complex geometry of accretion streams and the strong cyclotron radiation from accretion spots. Eclipses also help in reconstructing the geometry of accretion streams \citep{Harrop99} and the distribution of shock regions across the surface of the white dwarf \citep{Donoghue06}.

In this paper we present phase-resolved spectral and photometic study of the eclipsing polar SDSS J002637.06+242915.6 (also known as ZTF17aaaehby, Gaia 21ara, AT2021ceg; hereafter referred to as {\obj}). The object {\obj} was initially discovered as an optical transient by the Catalina Sky Survey \citep{Drake09}. Based on its high-amplitude photometric variations, detected by the Palomar Transient Factory (PTF) survey, and a strong HeII~$\lambda$4686 emission line, {\obj} was classified as a candidate to magnetic cataclysmic variables \citep{Margon14}. An analysis of Zwicky Transient Facility (ZTF) observations by \cite{Szkody24} revealed the orbital period of $P_{orb} \approx 122.9$~min and the presence of eclipses. The spectra exhibited a high radial velocity amplitude ($K>400$~km s$^{-1}$), characteristic of AM~Her-type stars.

\section{Observations and Data Reduction}
\subsection{Photometry}

Photometric observations of {\obj} were obtained with the 1-m Zeiss-1000 telescope at the Special Astrophysical Observatory of the Russian Academy of Sciences (SAO RAS). The telescope was equipped with an MMPP photometer\footnote{See more about MMPP photometer at https://www.sao.ru/Doc-k8/Telescopes/small/MMPP/\#1} with an Andor iXon Ultra 888 EMCCD detector ($1K \times 1K$ pixels).
Observations were made without photometric filters with a 10-sec exposures. A total of 720 exposures were obtained, covering the orbital period of the polar.
We applied the standard data reduction procedure and performed aperture photometry of {\obj} using the photutils library\footnote{The photutils library is available at https://photutils.readthedocs.io/en/stable/}.

\subsection{Spectroscopy}

Spectroscopic observations of {\obj} were performed on November 27/28, 2024 with the 6-m BTA telescope at SAO RAS using the focal reducer SCORPIO\footnote{A detailed description of the SCORPIO focal reducer can be found on https://www.sao.ru/hq/lsfvo/devices/scorpio/scorpio.html} in long-slit spectroscopy mode \citep{Afan05}. The instrument was equipped with a volume phase holographic grating VPHG550G (550 lines per mm), which with a  $1.2''$ slit width provided spectral coverage of $3800-7300$~\AA\, at a resolution of $\Delta \lambda \approx 12$~\AA. A total of 21 spectra with 300-sec exposures were obtained, covering approximately $0.9$ of the polar's orbital period. The observations were carried out in good astroclimatic conditions with a seeing of $\approx 1''$.

Data reduction followed standard procedures for long-slit spectroscopy, including cosmic rays removal, bias subtraction, flat-fielding, and wavelength calibration based on He-Ne-Ar lamp frames. The spectra were optimally extracted \citep{Horne86} with sky background subtraction. Spectrophotometric calibration was performed by the observations of the standard star BD+28$^{\circ}$~4211 \citep{Bohlin01}. We calculated the barycentric Julian dates (BJD) and barycentric radial velocity corrections for each spectrum.

\section{Photometric analysis}
\label{sec:phot}
\begin{figure}
\center{\includegraphics[width=1.0\textwidth]{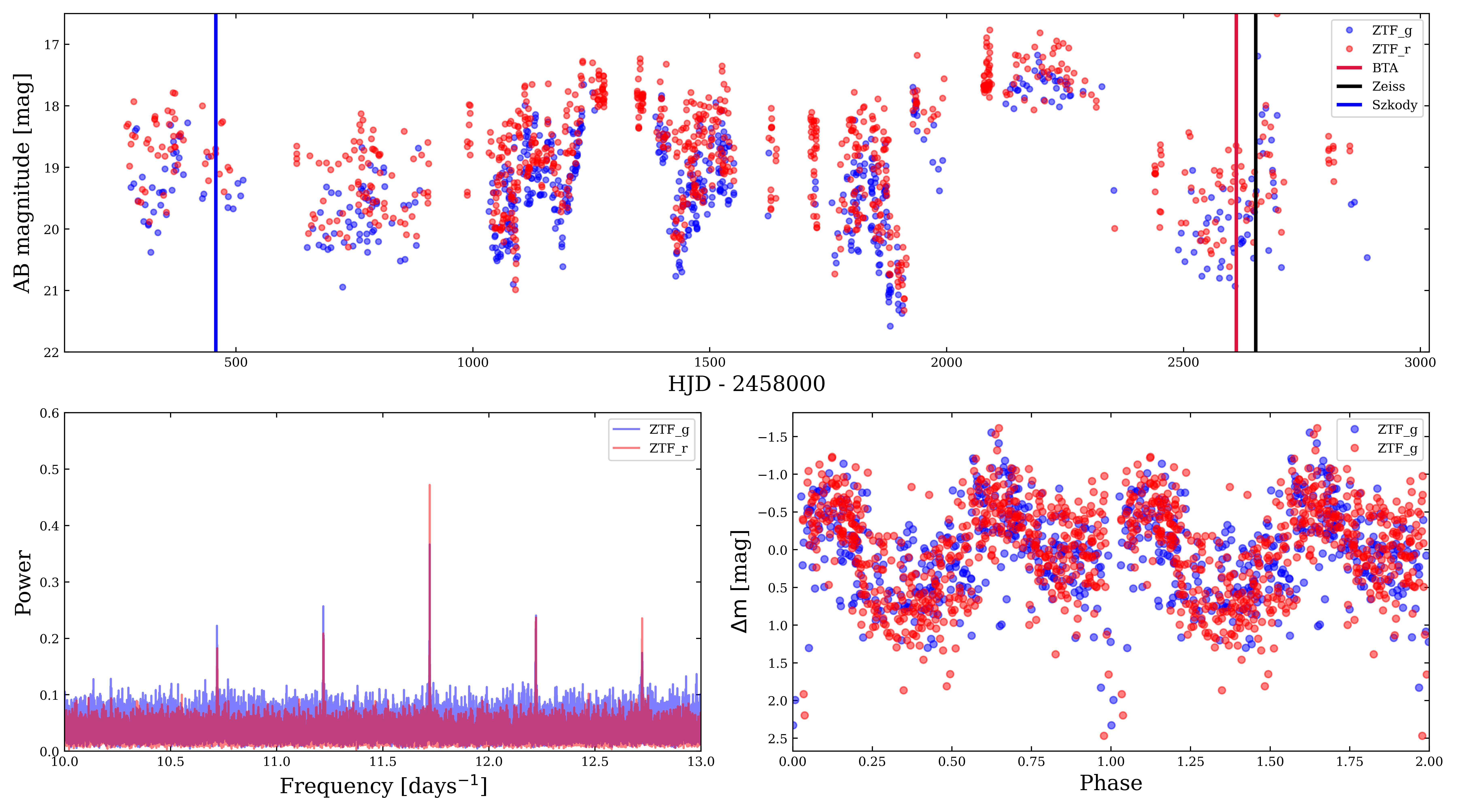}}
\caption{Top panel: long-term light curve of {\obj} from ZTF data in $g$ and $r$ bands.
Bottom left panel: Lomb-Scargle periodograms constructed from $g$- and $r$-band data.
Bottom right panel: phase-folded light curves in $g$ and $r$ bands. Vertical lines mark observation epochs from \cite{Szkody24} (blue line), BTA telescope (red), and Zeiss-1000 telescope (black)}
\label{fig:lc_ztf}
\end{figure}

Figure~\ref{fig:lc_ztf} shows the light curve of {\obj} obtained by ZTF (\citealt{Masci19}) survey in the $g$ and $r$ bands. The eclipses were excised from the light curve. 
The long-term light curve shows clear transitions between a low state ($\langle r \rangle \approx 20.5^m$) and a high state ($\langle r \rangle \approx 17.5^m$).
Using data close to the intermediate state ($\langle r \rangle \approx 19^m$), the orbital period $P_{orb} = 122.856 \pm 0.006$~min was determined by the Lomb-Scargle method (the error was measured by the Monte Carlo method and corresponds to the $1\sigma$ level). The ephemeris of eclipse midtimes was obtained:
\begin{equation}
\mathrm{HJD}_{\min} = 2458000.06(2) + 0.085316(4) \times E,
\end{equation}
where $E$ is the orbital cycle number. The Lomb-Scargle periodograms and phase light curves of {\obj} in the intermediate state are shown in Fig.~\ref{fig:lc_ztf}. One can see that the light curves exhibit double-peaked out-of-eclipse variability with amplitudes $\Delta g \approx \Delta r \approx 1.5^m$.

\begin{figure}
\centering
\includegraphics[width=1\textwidth]{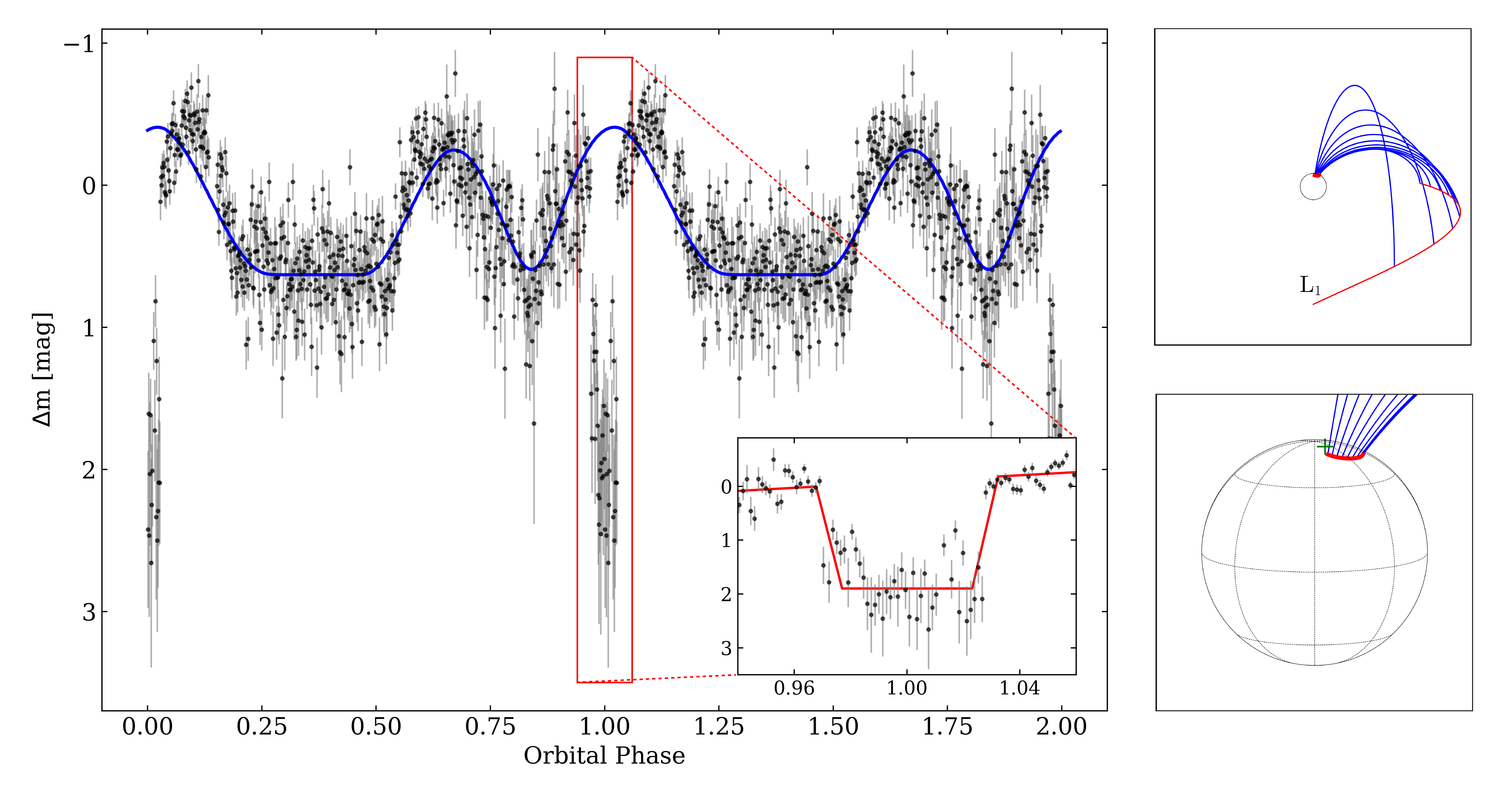}
\caption{Left panel: phase-folded light curve of {\obj} obtained with the Zeiss-1000 telescope (black points with gray error bars). The blue curve shows the model light curve of the accreting white dwarf. The inset illustrates the trapezoidal fit to the eclipse profile.
Right panels: accretion model for {\obj} as viewed by the observer at phase $\varphi=0$, shown in two different scales. The thin red line indicates the ballistic stream trajectory, while blue lines represent the magnetic trajectory. The accretion spot on the white dwarf is marked by a thick red line, and the magnetic pole position is denoted by a green cross.}
\label{fig:lc_zeiss}
\end{figure}

The light curve of {\obj} obtained with the Zeiss-1000 telescope is shown in Fig.~\ref{fig:lc_zeiss}. It exhibits a trapezoidal eclipse with a depth of $\approx 2^m$. By fitting the eclipse profile with a trapezoid, we determined the eclipse width at half depth $\Delta t_{ecl} = 448.6\pm 1.1$~s and the ingress duration (equal to egress duration) $\Delta t_{ing} = 67.9 \pm 1.5$~s. The light curve reveals a double-peaked M-shaped bright phase ($\varphi \in 0.55-1.20$). This phenomenon is common in polars and is interpreted as cyclotron beaming, meaning the radiation intensity strongly depends on the angle $\theta$ between the magnetic field line direction and the line of sight. Under certain conditions, cyclotron radiation intensity peaks at $\theta = 90^{\circ}$. For high white dwarf rotation axis inclination ($i\approx 90^{\circ}$), the cyclotron flux maximum occurs when the spot is near the stellar limb ($\theta \approx 90^{\circ}$) and reaches a minimum at the spot's smallest distance to the center of the stellar disk (where $\theta$ is smallest). This produces the M-shaped profile in the light curve, covering approximately half of the white dwarf's rotation period (see, e.g., \citealt{Kolbin20}). The light curve additionally exhibits a plateau phase ($\varphi \in 0.20-0.55$), likely corresponding to the accretion spot being occulted behind the white dwarf's disk. For comparison, we overplot a theoretical light curve computed using the code from \cite{Kolbin20}. The simulation assumed an accretion spot with a magnetic field strength $B=34$~MG and a temperature $T = 18$~keV, which are consistent with the cyclotron spectra of {\obj} (see Section \ref{sec:cyc_spec}).
The magnetic dipole is inclined at $\beta = 11^{\circ}$ relative to the rotation axis, with the magnetic pole located at longitude $\psi = 31^{\circ}$ measured from the direction to the donor star. The mass ratio and orbital inclination were fixed at $q=0.3$ and $i=79^{\circ}$ respectively (see Section \ref{sec:pars}).
Figure~\ref{fig:lc_zeiss} shows the white dwarf accretion model that describes the light curve of {\obj}. However, due to significant uncertainties in the magnetic field strength and spot temperature parameters, the light curve modeling cannot provide reliable constraints on either the magnetic dipole orientation or the geometric parameters of the accretion spot.

\section{Spectral analysis} 

\subsection{Zeeman splitting}
\label{sec:zee}

The spectra of {\obj} can be divided into two distinct groups. The first group corresponds to the plateau phase of the light curve ($\varphi \in 0.20-0.55$). These spectra exhibit a blue continuum, characteristic of white dwarf. However, they show no photospheric absorption lines or any evidence of a cold donor in the red spectral region.
The second group of spectra was obtained during the bright phase, when the accretion spot becomes visible on the white dwarf's disk. These spectra show the red continuum produced by cyclotron radiation from the accretion spot. Such spectra are commonly observed in polars with relatively low magnetic fields ($B \lesssim 30$~MG; see \citealt{Schwope97c,Schwope24,Thomas96}).
Notably, near phase $\varphi\approx 0.8$, the cyclotron contribution becomes minimal, and the spectra show a blue slope. 

\begin{figure}
\center{\includegraphics[width=1\textwidth]{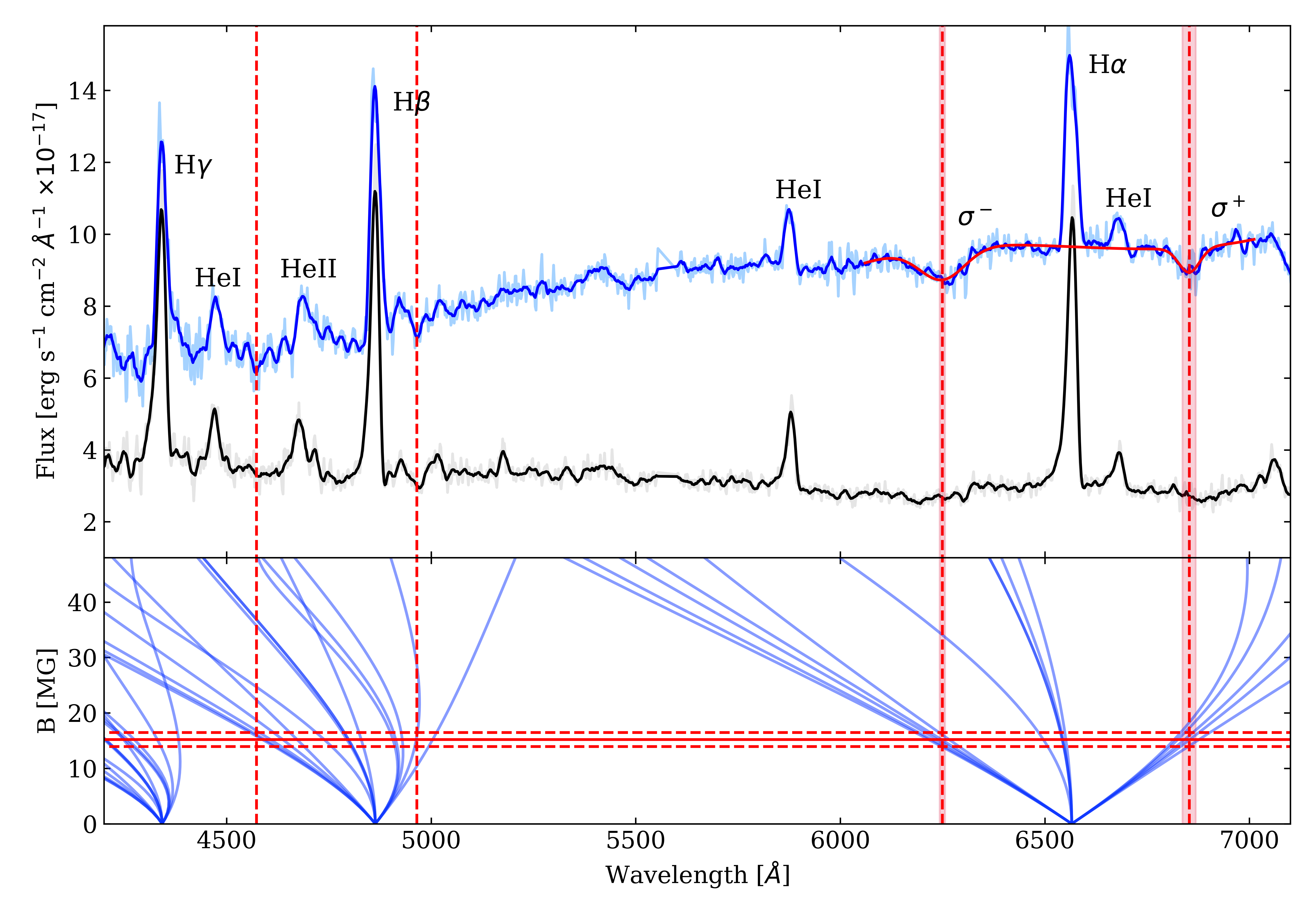}}
\caption{Top panel: averaged spectra of {\obj} for the bright phase peaks (blue line) and plateau phase (gray line). Blue and black lines show Savitzky-Golay filtered spectra. The red line represents the fit to the H$\alpha$ line region using a low-order polynomial combined with two Gaussian components. Vertical lines mark the positions of the H$\alpha$ Zeeman triplet components and likely H$\beta$ splitting components.
Bottom panel: Zeeman splitting diagrams for H$\alpha$, H$\beta$, and H$\gamma$ lines. Horizontal lines indicate the magnetic field estimate and its uncertainty.}
\label{fig:spec_zee}
\end{figure}

Near the brightness peaks (phases $\varphi \approx 0.6$ and $\varphi \approx 0.1$), the Zeeman absorption components of the H$\alpha$ line become distinct (see Fig.~\ref{fig:spec_zee}). The Zeeman splitting is observed only during the accretion spot's visibility phase on the white dwarf's disk, suggesting its origin in the cold halo above the accretion spot (similar phenomena are described in \citealt{Kolbin23}). Comparison of the $\sigma^-$ and $\sigma^+$ component positions with the splitting diagram yields a magnetic field strength of $B = 15.1 \pm 1.3$~MG (see Fig.~\ref{fig:spec_zee}). The $\sigma^+$ component blends with the telluric Fraunhofer B-band, requiring division of the spectra of {\obj} by continuum-normalized standard star spectra for isolation. The Zeeman components show no detectable shifts within $\Delta \lambda \approx 17$~\AA, corresponding to a magnetic field variation of $\Delta B = 3.1$~MG. The splitting diagram was calculated using the \cite{Schimeczek14} code.
Weak flux dips near 4570~\AA\ and 4965~\AA, comparable to the noise level, are present in the bright-phase spectra. These likely correspond to the Zeeman components of the H$\beta$ line.

\subsection{Emission lines}
\label{lines}

The obtained spectra of {\obj} exhibit emission lines typical of cataclysmic variables, including hydrogen Balmer lines, neutral helium lines, and the ionized helium line HeII~$\lambda4686$. The HeII~$\lambda4686$ line is less intense than in the spectra presented by \cite{Margon14} and \cite{Szkody24}, where its strength was comparable to that of H$\beta$. This suggests that our observations occurred during a state of reduced accretion rate, which is consistent with the fainter average magnitude we observed ($\langle r \rangle \approx 19.2^m$) compared to the $\langle r \rangle \approx 18.8^m$ reported by \cite{Szkody24} (see also Fig.~\ref{fig:lc_ztf})

\begin{figure}
\center{\includegraphics[width=1\textwidth]{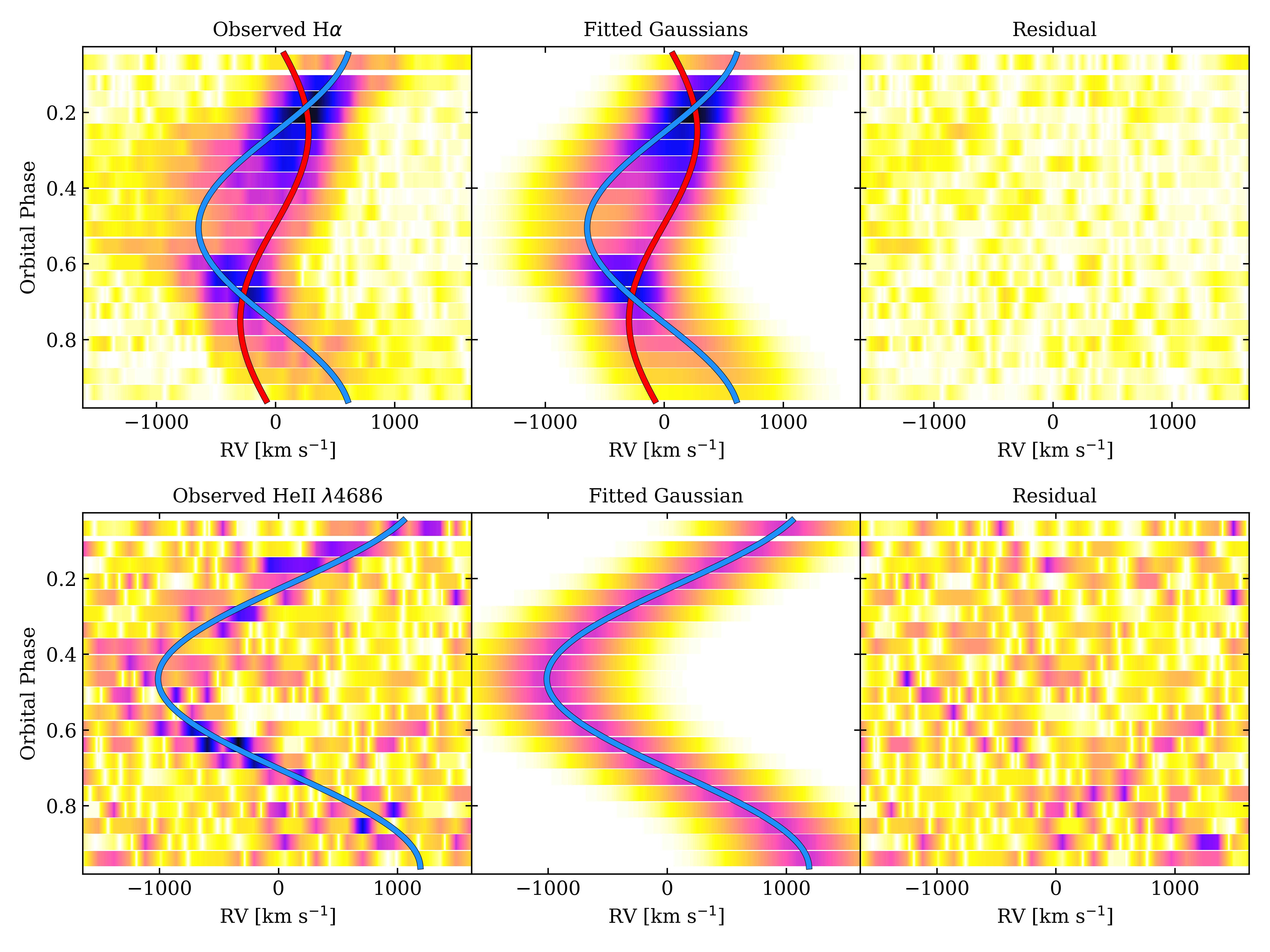}}
\caption{Top panel: H$\alpha$ trailed spectra (left), sum of fitted Gaussian components (middle), and residual spectrum (right). Radial velocity curves of the narrow (red) and broad (blue) components are overplotted.
Bottom panel: HeII~$\lambda4686$ trailed spectra (left), fitted Gaussian (middle), and residual spectrum (right).}
\label{fig:spec_din}
\end{figure}

The orbital variability of the H$\alpha$ and HeII~$\lambda4686$ lines is shown in the trailed spectra presented in Fig.~\ref{fig:spec_din}. Other Balmer lines have behavior similar to the H$\alpha$ line, but have a lower signal-to-noise ratio. The HeII~$\lambda4686$ line displays a larger radial velocity semi-amplitude ($\approx1000$ km s$^{-1}$).
In polars, spectral line profiles are often dominated by a narrow and a broad components (\citealt{Liu23, Kolbin23, Lin25}). We analyzed the H$\alpha$ orbital variability by decomposing its profile $f(v)$ into two Gaussians:
\begin{equation}
f(v) = B + \sum_{i\in{n,b}}A_i\exp\left[-\frac{(v-V_i)^2}{2\sigma_i^2}\right],
\label{gauss}
\end{equation}
where $v$ is the radial velocity at a given profile point, $A$ is the Gaussian amplitude, $\sigma$ is the standard deviation, $V$ is the Gaussian center velocity, and $B$ is the continuum level near the line. The indices $n$ and $b$ correspond to the narrow and broad components, respectively. The motion of spectral profile components was modeled as sinusoidal:
\begin{equation}
V_i=\gamma + K_i\sin{ \big[2\pi (\varphi-\varphi^0_{i})\big]},
\label{eq_sin}
\end{equation}
where $i \in {n,b}$, $\gamma$ is the system's center-of-mass radial velocity, $K$ is the radial velocity semi-amplitude, $\varphi$ is the orbital phase, and $\varphi^0$ is the initial phase.

The parameters $\gamma$, $K_{n,b}$, and $\varphi^0_{n,b}$ were derived through least-squares fitting of the trailed spectra. The differential evolution algorithm was used to minimize $\chi^2$. The Gaussian amplitudes $A_{n,b}$ and continuum level $B$ were determined individually for each profile, while the $\sigma_{n,b}$ parameters (assumed time-independent) together with $\gamma$, $K_{n,b}$, and $\varphi^0_{n,b}$ were found by fitting the entire set of profiles. The parameter uncertainties were estimated via Monte Carlo simulations.
The reconstructed radial velocities for both components are shown in Fig.~\ref{fig:spec_din}. Semi-amplitudes of $K_n = 293 \pm 48$ km s$^{-1}$ and $K_b = 814 \pm 140$ km s$^{-1}$ were measured for the narrow ($\sigma\approx230$ km s$^{-1}$) and broad ($\sigma\approx350$ km s$^{-1}$) components, respectively.
The initial phase for the narrow component is $\varphi^0_{n} = 0.08\pm0.03$, while for the broad component $\varphi^0_{b} = 0.23\pm0.06$. 

We suggest that, as in some other polars, the narrow component may form on the donor's surface X-ray heating. The radial velocities of the narrow component generally agree with the orbital motion of the irradiated hemisphere of the donor. The trailed spectra show no trace of the narrow component near phase $\varphi=0$. This may occur when the irradiated region passes behind the donor's disk, supporting our interpretation. The small deviation from $\varphi^0_n = 0$ might be caused by non-uniform irradiation, possibly due to shielding by the accretion stream \citep{Schwope97}. The trailed spectra show increased flux in the emission line around phases $\varphi \approx 0.2$ and $\varphi \approx 0.7$, a phenomenon commonly observed in AM~Her systems and interpreted as the projection effect of the optically thick accretion stream (\citealt{Simic98,Silber92}).

\subsection{Doppler tomography}

The Doppler tomography of {\obj} was performed to identify the regions of emission line formation. This method converts the trailed spectra of emission lines into the map of the emission sources in two-dimensional velocity-space (see, e.g., \citealt{Marsh05}). In the velocity space, positions are often described by two polar coordinates: the absolute velocity relative to the binary's center of mass $\upsilon$ (scaled by $\sin i$) and the angle $\theta$ between the particle's velocity vector and the line connecting the white dwarf and donor centers (see \citealt{Kotze15} for details). The reconstruction of tomograms was carried out using the code of \cite{Kotze15} implementing maximum-entropy method.

\begin{figure}
\center{\includegraphics[width=10cm]{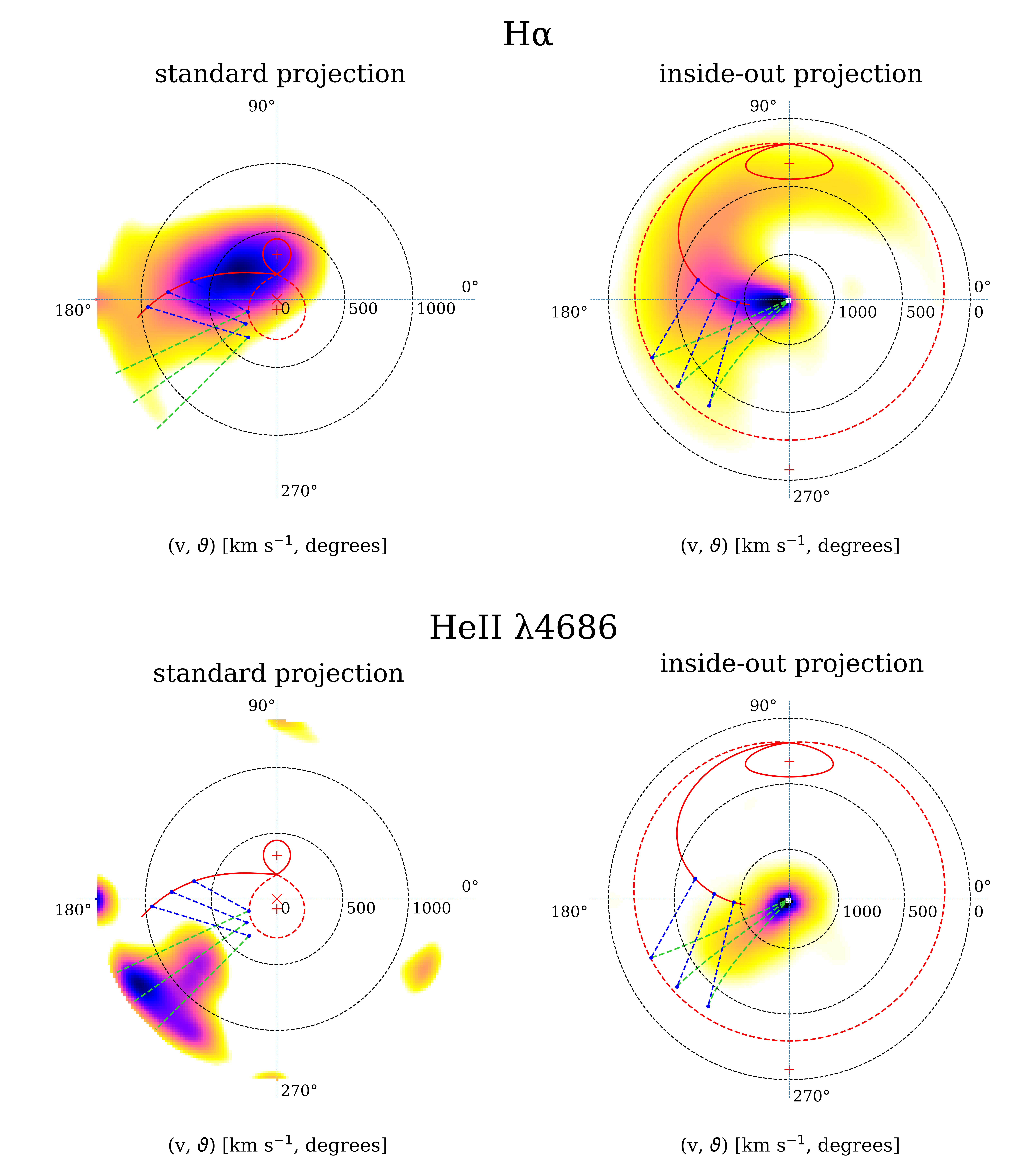}}
\caption{Doppler tomograms of {\obj} in the H$\alpha$ line (top) and HeII $\lambda$4686 (bottom), shown in standard (left) and inside-out (right) projections. The tomograms are overlaid with the donor star's velocity (closed red curve) and the accretor's Roche lobe velocity. The open red curve marks the ballistic stream trajectory, while green curves correspond to particle velocities along the magnetic trajectory.}
\label{fig:dopptom_Ha_He}
\end{figure}

Figure \ref{fig:dopptom_Ha_He} shows Doppler tomograms of {\obj} in the H$\alpha$ and HeII~$\lambda4686$ lines, presented in the standard and inside-out projections. In the first projection, the absolute velocity $v$ increases from the center of the tomogram to the periphery, while in the second, it increases from the periphery to the center. The latter variant is convenient for studying high-velocity regions of the accretion stream, which are heavily smeared out on maps in the standard projection (for more details on this effect, see \citealt{Kotze15}).
Although the obtained tomograms are rather smeared due to the low spectral resolution ($\Delta \lambda \approx 12$~\AA), they clearly show the accretion stream feature typical for polars. For the interpretation of the tomograms, the system model used in Section \ref{sec:phot} to describe the light curve (see also Fig.~\ref{fig:lc_zeiss}) is overplotted.
On the tomograms in the H$\alpha$ line, the emission regions are located near the ballistic trajectory. A different pattern is observed for the HeII~$\lambda4686$ line. The position of the emission source is in good agreement with the magnetic part of the stream trajectory, while there are no signs of emission on the ballistic trajectory. It can be assumed that the magnetic part of the stream effectively absorbs the hard ultraviolet radiation from the accretion spot, preventing it from reaching the ballistic part of the stream and the surface of the donor star \citep{Schwope97b}.

\begin{figure}
\center{\includegraphics[width=\textwidth]{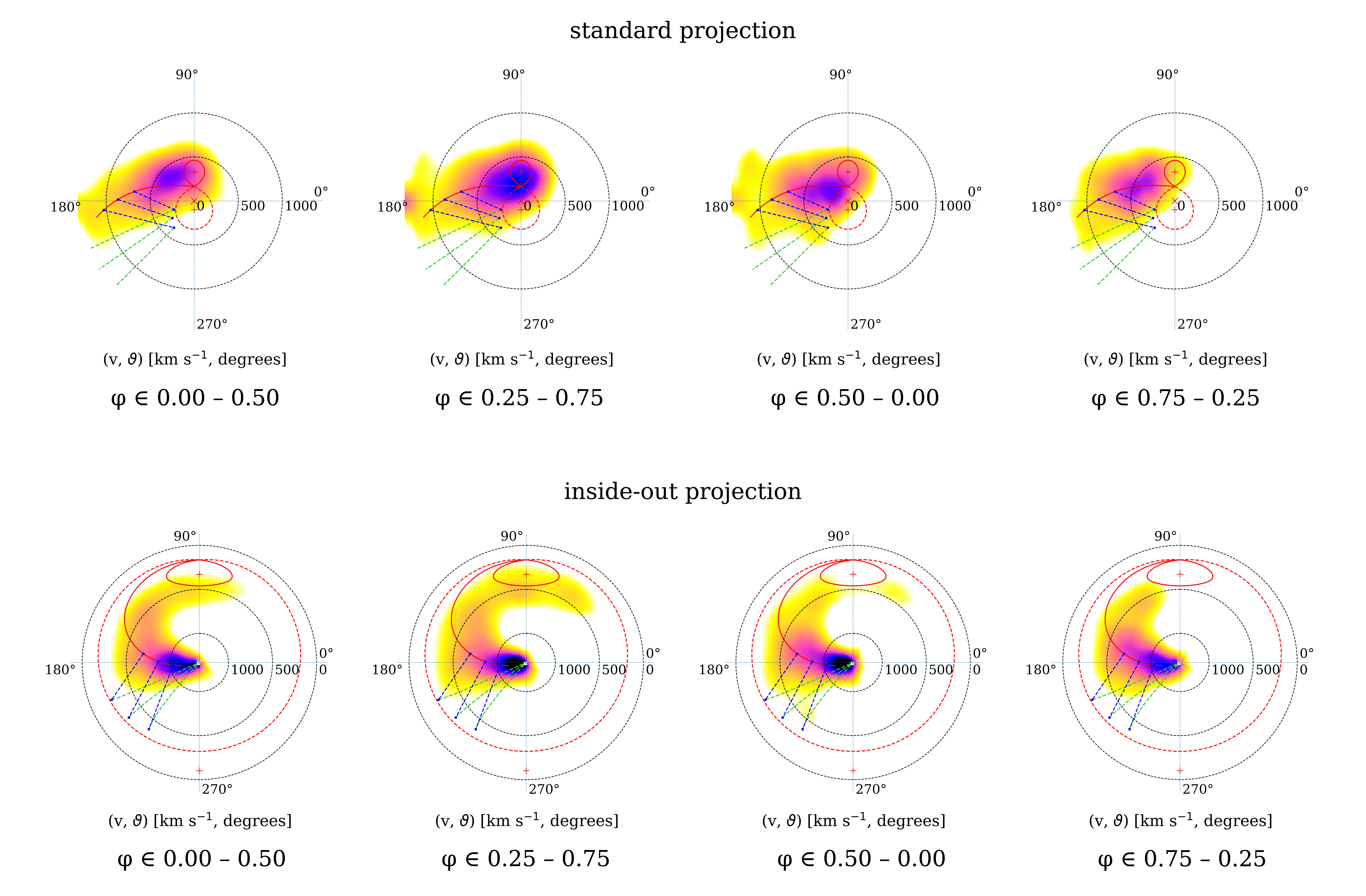}}
\caption{Half-phase Doppler tomograms of {\obj} for the H$\alpha$ line in the standard (top) and inside-out (bottom) projections. The ranges of orbital phases used to construct the tomograms are indicated.}
\label{fig:dopptom_out_Ha}
\end{figure}

Figure \ref{fig:dopptom_out_Ha} shows half-phase Doppler tomograms constructed from sections of the trailed spectra with a width of half the orbital period. Such tomograms allow us to determine the phases of best visibility for different regions of the system. The tomograms were obtained in the H$\alpha$ line and are presented in both standard and inside-out projections.
Throughout the entire cycle, a high-velocity region ($v \gtrsim 500$~km s$^{-1}$, $\vartheta \approx 180^{\circ}$) is observed, likely located on the far side of the ballistic trajectory. Variability is noticeable in the emission near $\vartheta =90^{\circ}$. The emission in this region is maximum around phase $\varphi \approx 0.5$ and is not present around phase $\varphi \approx 0$. This behavior is expected for the irradiated hemisphere of the donor star, which has the best visibility  around phase $\varphi = 0.5$ and is hidden behind the donor's disk at phase $\varphi = 0$.

\subsection{Cyclotron spectra}
\label{sec:cyc_spec}
\begin{figure}
\center{\includegraphics[width=8cm]{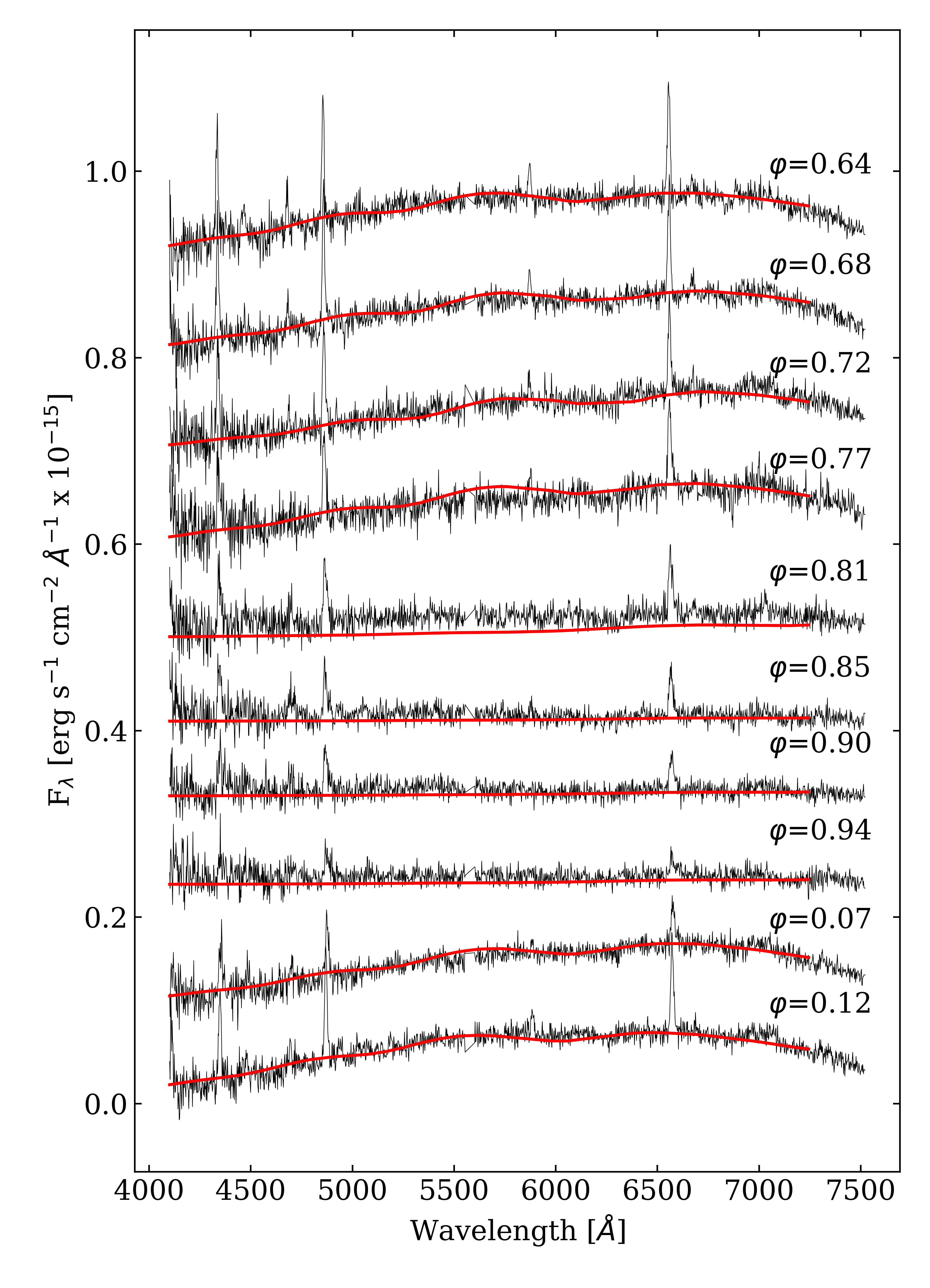}}
\caption{Cyclotron spectra of {\obj} in the bright phase (black line) and the model cyclotron spectra (red line). For clarity, the spectra are shifted along the ordinate axis relative to the spectrum at phase $\varphi=0.12$ by $\Delta F_\lambda\approx1\times10^{-16}$ erg s$^{-1}$ cm$^{-2}$ \AA$^{-1}$.}
\label{fig:cyc_spec}
\end{figure}

As noted in Section \ref{sec:zee}, the bright phase spectra exhibit a red cyclotron continuum without pronounced cyclotron harmonics. It is possible to impose constraints on magnetic strength $B$ and the mean electron temperature of the accretion spot $T_e$ by modeling a set of these spectra. The cyclotron spectra of {\obj} are shown in Fig.~\ref{fig:cyc_spec}. They were obtained by subtracting the averaged spectrum of the plateau phase from the bright phase spectra. The plateau spectrum is assumed to include only radiation from the white dwarf and the accretion stream. A simple model of an accretion spot, uniform in temperature and density, was used to calculate the theoretical cyclotron spectra. This model is often employed to estimate accretion spot parameters in polars (see, for example, \citealt{Kolbin22, Lin25, Beuermann21}). Within this model, the emergent intensities for the ordinary ($o$) and extraordinary ($e$) modes are defined as
\begin{equation}
I_{o,e} =\frac{I_{RJ}}{2}[1-\exp(-\alpha_{o,e}\Lambda)],
\label{eq_con}
\end{equation}
where $I_{RJ}/2=k_B T_e\omega^2/16\pi^{3}c^{2}$ is the Rayleigh-Jeans intensity per polarization mode ($k_B$ is the Boltzmann constant, $c$ is the speed of light), $\alpha_{o,e}$ are the cyclotron absorption coefficients in units of $\omega^2_p / \omega_c c$ for ordinary and extraordinary waves ($\omega_{p}$ is the plasma frequency, $\omega_c$ is the cyclotron frequency), and $\Lambda= \omega^{2}_{p} \ell/\omega_{c}c$ is the plasma parameter, which depends on the thickness of the emitting region along the line of sight $\ell$. The absorption coefficients $\alpha_{o,e}$ depend on the frequencies ratio $\omega/\omega_c$, the temperature of the emitting medium $T_e$, and the angle $\theta$ between the magnetic field lines and the line of sight. They were calculated according to the method described in~\cite{Chan81}. The total intensity of the cyclotron radiation is defined as the sum of the intensities of the ordinary and extraordinary waves, i.e., $I_{cyc} = I_o + I_e$.
The shape of the cyclotron spectrum depends on the magnetic strength $B$, the electron temperature $T_{e}$, the plasma parameter $\Lambda$, and the angle $\theta$ between the magnetic field lines and the line of sight.
The modeling of the entire set of cyclotron spectra was performed with unified values of $B$ and $T_e$, while the parameters $\theta$ and $\Lambda$ were determined individually for each spectrum (for more details, see \citealt{Kolbin23}).

\begin{figure}
\center{\includegraphics[width=8cm]{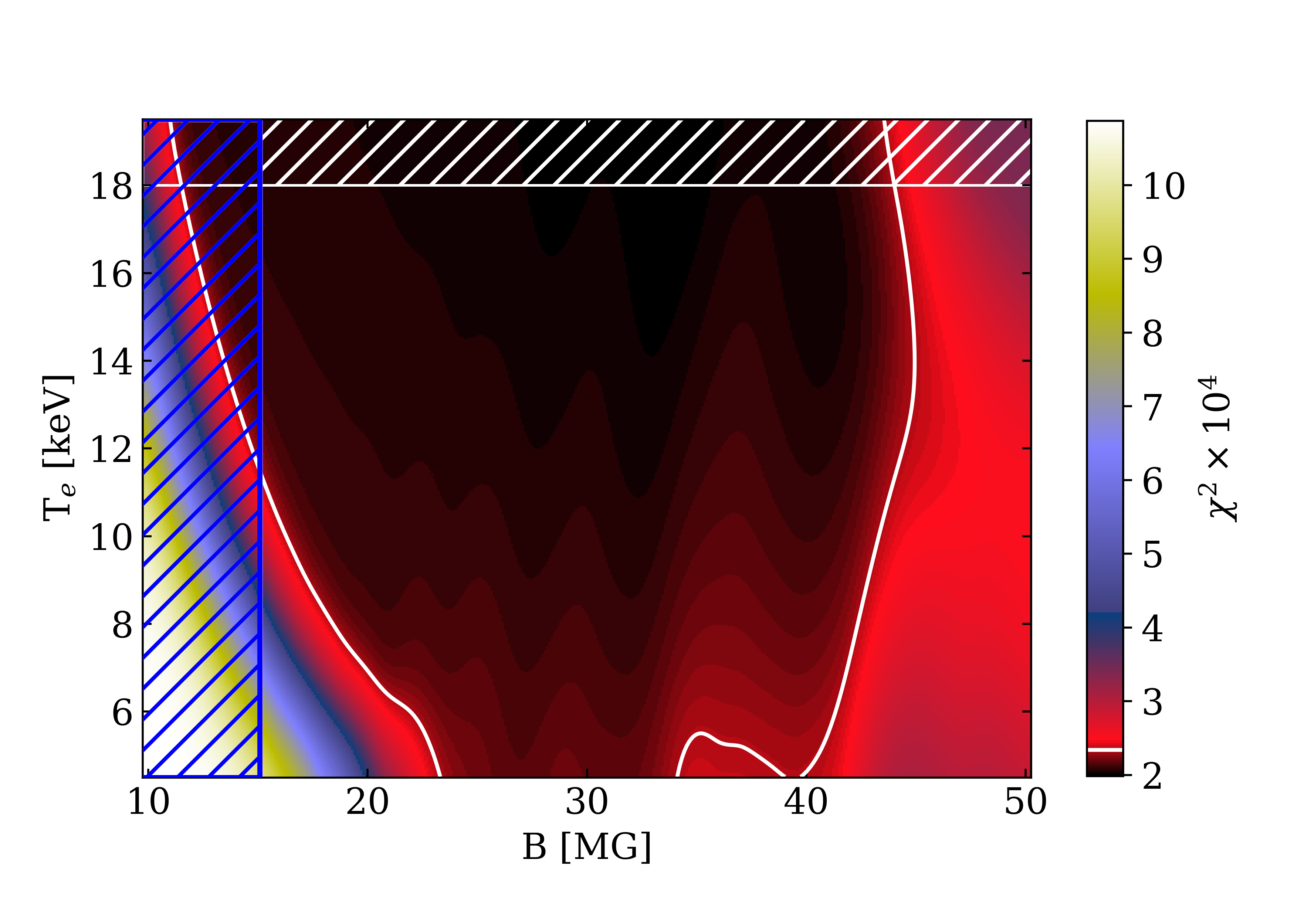}}
\caption{Maps of the minimum $\chi^2$ distribution in the $B - T_e$ plane. The white contour indicates the 1$\sigma$ confidence level. The area hatched in blue denotes the range of magnetic field values below the estimate obtained from Zeeman splitting, $B_z=15.1$~MG. The area hatched in white indicates the region of temperatures exceeding the maximum post-shock temperature $T_{sh}=18$~keV.}
\label{fig:map_BT}
\end{figure}

Due to the absence of pronounced cyclotron harmonics, we obtained a satisfactory description of the spectra over a wide range of parameters. Figure \ref{fig:map_BT} shows a map of the minimum $\chi^2$ distribution in the $B$-$T_e$ plane, along with the $\chi^2-\chi^2_{\min}$ contour corresponding to the 1$\sigma$ (or 68\%) confidence level (see, for example, \citealt{Baron13}). The best description of the spectra is shown in Fig.~\ref{fig:cyc_spec} and was obtained at a magnetic field of $B = 34$~MG and a temperature of $T_e = 18$~keV.
The solution allows for a magnetic strength in the accretion spot  up to $B_{cyc} \lesssim 45$~MG. The magnetic field estimate of $B \approx 15.1$~MG, obtained from Zeeman splitting, can be considered a lower limit. The post-shock temperature \citep{Aizu73} can be adopted as an upper temperature limit:
\begin{equation}
T_{sh} = \frac{3}{8}\frac{G M_1 m_H \mu}{k_B R_1},
\label{eq_Tsh}
\end{equation}
where $G$ is the gravitational constant, $M_1$ and $R_1$ are the mass and radius of the white dwarf, respectively, $m_H$ is the mass of the hydrogen atom, and $\mu = 0.615$ is the mean molecular weight for solar composition. For the white dwarf {\obj} with a mass $0.42 \le M_1/M_\odot \le 0.72$ (see Section~\ref{sec:pars}), the post-shock temperature is $T_{sh}\approx8-18$~keV. 

\section{Binary parameters}
\label{sec:pars}
\begin{figure}
\center{\includegraphics[width=8cm]{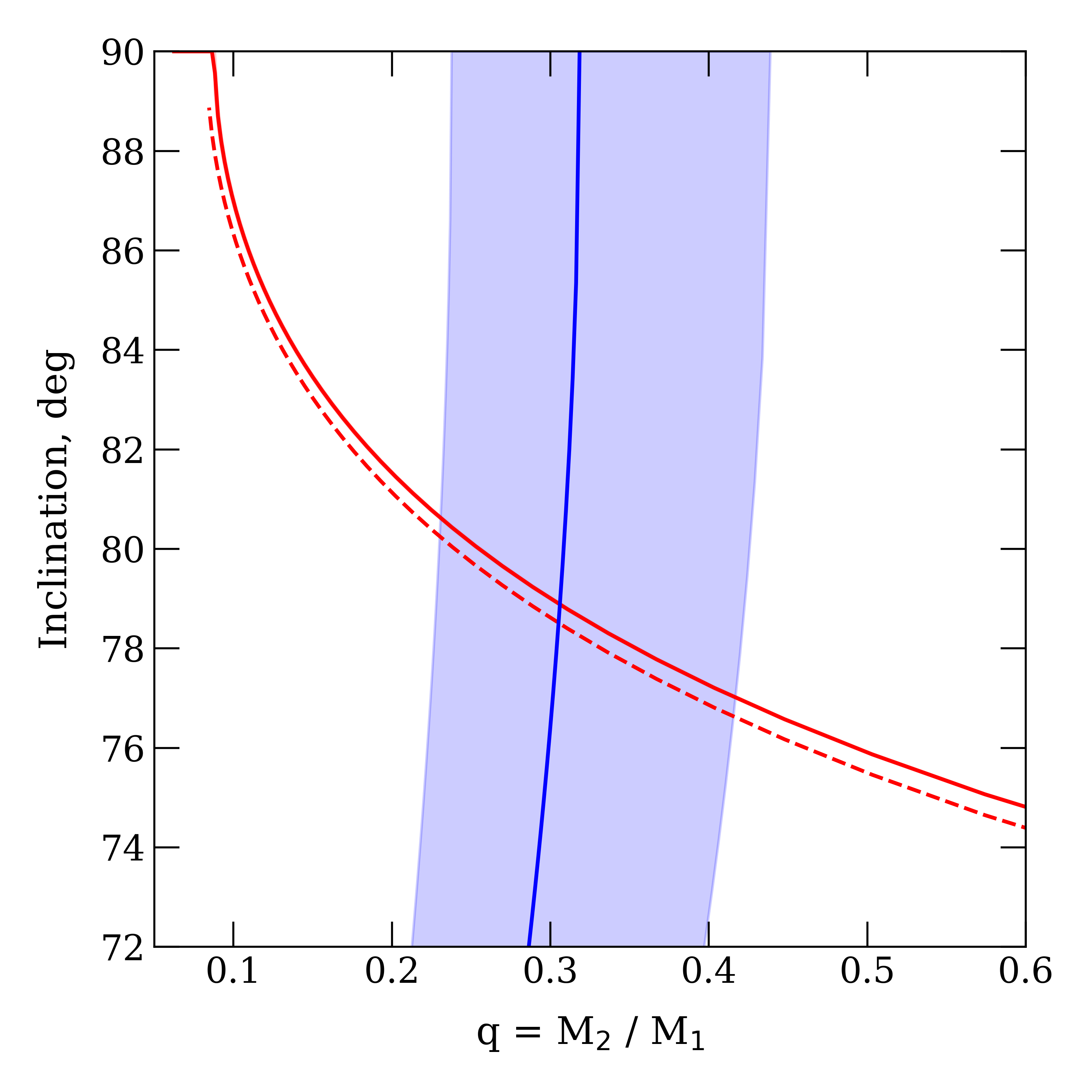}}
\caption{Set of solutions in the $q-i$ plane. The solid red line shows the solution that provides the observed eclipse duration if it was caused by the occultation of an unspotted white dwarf by the donor. The dashed red line corresponds to the maximum deviation from the first solution caused by the eclipse of a bright accretion spot. The solid blue line is the solution that reproduces the donor's radial velocity curve in the H$\alpha$ line.
}
\label{fig:qi}
\end{figure}

The eclipse duration of the white dwarf in cataclysmic variables is a function of the orbital inclination $i$ and the mass ratio $q=M_2/M_1$ \citep{Horne85}. We constructed a model of a semi-detached binary system and used it to find a solution in the $q$–$i$ plane that reproduces the observed eclipse duration in {\obj} of $\Delta t_{ecl} = 448.6\pm 1.1$~s. This solution is shown in Fig.~\ref{fig:qi}. It can be seen that it places a constraint on the mass ratio of $q\ge 0.086$. We note that in addition to the white dwarf, a bright accretion spot is also eclipsed in {\obj}. The accretion spot shifts the centre of light towards the donor and slightly broadens the eclipse \citep{Schwope97}. Figure \ref{fig:qi} shows the maximum possible shift of the solution, obtained by modelling the eclipse of the point on the white dwarf surface closest to the donor (the white dwarf radius was assumed to be $R_1 = 0.013 R_\odot$; see below). It is evident that the accretion spot can cause an overestimation of the mass ratio by $\Delta q \approx 0.02$ and of the orbital inclination by $\Delta i \approx 0.4^\circ$.

Another constraint on the binary parameters can be derived from the radial velocities of the irradiated hemisphere of the donor. In Section \ref{lines}, we obtained an estimate for the semi-amplitude of the radial velocity of the irradiated hemisphere of $K_n = 293\pm48$~km s$^{-1}$. The semi-amplitude of the radial velocity of the donor's center of mass can be expressed as $K_2 = K_n + \Delta K$, where $\Delta K$ is a correction that can be estimated theoretically. To determine $\Delta K$, we used the code of \cite{Shimanskii12}. This code solves the radiative transfer equation for irradiated stellar atmospheres and performs spectrum synthesis for semi-detached binary systems, accounting for irradiation effects. Using this code, we generated a set of radial velocity curves for the H$\alpha$ emission lines of semi-detached binaries with an externally irradiated donor. These curves were then approximated with sinusoids to determine the corrections $\Delta K$. This method yielded a correction of $\Delta K = 36$~km s$^{-1}$. The corresponding mass function, $f_2(m) = P_{orb} K^3_2 / 2\pi G = 0.32 \pm 0.14~M_\odot$, provides a lower limit on the mass of the white dwarf. A solution in the $q$–$i$ plane can be found using the definition of the mass function:
\begin{equation}
f_2(m) = \frac{M_2 \sin^3{i}}{q(1 + q)^2}.
\label{fm}
\end{equation}
The ambiguity in the donor mass $M_2$ can be eliminated by requiring the equality of the effective radius of the critical Roche lobe $R_L$ and the evolutionary radius $R_2$, which is related to the mass $M_2$, i.e.,
\begin{equation}
R_L(q,M_2) = R_2(M_2).
\end{equation}
Since {\obj} is located below the period gap, its donor must be fully convective \citep{Knigge11}. Therefore, the effective radius of the critical Roche lobe can be calculated using the formula
\begin{equation}
R_L=A\frac{0.5126 q^{0.7388}}{0.6710q^{0.7349}+\mathrm{ln}(1+q^{0.3983})},
\label{eq_rl}
\end{equation}
derived by \cite{Sirotkin09} for polytropic models with an index of $n=1.5$ (which corresponds to fully convective stars). The semi-major axis $A$ is found from Kepler's third law: $A=(M_2(1+1/q)P_{orb}^2)^{1/3}$.
For the evolutionary dependence $R_2(M_2)$, we used the relation
\begin{equation}
\frac{R_2}{R_{\odot}} = 0.225 \pm 0.008 \left( \frac{M_2}{M_{\mathrm{conv}}} \right)^{0.636 \pm 0.012},
\label{eq_r2}
\end{equation}
obtained by \cite{McAllister19}, where $M_{\mathrm{conv}} = 0.20 \pm 0.02~M_{\odot}$. This relation is valid for donors in non-magnetic cataclysmic variables within the mass range $M_2 \in M_{\mathrm{bounce}} - M_{\mathrm{conv}}$ ($M_{\mathrm{bounce}} = 0.063~M_{\odot}$), which have evolved through the period gap but have not yet become period bouncers.

The solution in the $q-i$ plane corresponding to the donor mass function is shown in Fig.~\ref{fig:qi}. It can be seen that it is consistent with the eclipse duration at $i=78.9\pm1.7^{\circ}$ and $q=0.33\pm0.10$. From the equality of the evolutionary relation $R(M_2)$ and the effective Roche lobe radius, we find the donor mass $M_2 = 0.174\pm0.022 M_{\odot}$ and the corresponding radius $R_2 = 0.206\pm0.009 R_{\odot}$. Now, knowing the donor mass and the mass ratio, we obtain the white dwarf mass $M_1 = 0.57\pm 0.15 M_{\odot}$. The corresponding radius, according to the mass-radius relation for white dwarfs \citep{Nauenberg72}, is $R_1 = 0.013\pm0.002 R_{\odot}$. The parameter uncertainties were determined using the Monte Carlo method.

\section{Conclusion}

This work presents a spectral and photometric study of the eclipsing polar {\obj}. The main parameters of the investigated object, determined in this work, are given in Table~\ref{tab:all_pars}. Analysis of the long-term light curve from ZTF data revealed high and low states, differing in mean brightness by $\Delta r \approx 3^m$. We have refined the orbital period of the system to $P_{orb} = 122.856\pm0.006$~min. Based on observations with the Zeiss-1000 telescope, an updated value for the eclipse duration of $\Delta t_{ecl} = 448.6 \pm 1.1$ s was obtained. The light curve has an M-shaped bright phase, which has been interpreted by a simple accreting white dwarf model.

The spectra of the bright phase exhibit a Zeeman absorption triplet of the H$\alpha$ line superimposed on a red cyclotron continuum. The absorption triplet apparently forms in a cold halo surrounding the accretion spot. The magnetic field induction in the halo was estimated at $B=15.1\pm 1.3$ MG. Unfortunately, the absence of cyclotron harmonics does not allow us to determine the magnetic strength in the accretion spot and we only imposed constraints in a wide range: $15~\text{MG} \lesssim B_{cyc} \lesssim 45~\text{MG}$. In terms of spectral properties {\obj} resembles the polars EP~Dra \citep{Schwope97b}, BM~CrB \citep{Kolbin23}, and Gaia~23cer \citep{Kolbin24}.

The Doppler tomography indicates a difference in the regions of formation of the H$\alpha$ and HeII~$\lambda4686$ emissions. The H$\alpha$ line is formed on/near the ballistic trajectory of the accretion stream and on the irradiated hemisphere on the donor, while the orbital variability of the ionized helium line is consistent with its formation on the magnetic trajectory. The agreement in the orientation of the magnetic dipole obtained from the light curve modeling with the Doppler maps is noteworthy.

In the H$\alpha$ line profiles, a narrow component with a radial velocity half-amplitude of $K_n = 293$ km s$^{-1}$ appears. Analysis of the Doppler tomograms indicates that it originates on the irradiated hemisphere of the donor. Constraints on the orbital inclination, $77.2^\circ \le i \le 80.6^\circ$, the mass ratio, $0.23 \le q \le 0.43$, and the white dwarf mass, $0.42 \le M_1/M_\odot \le 0.72$, were derived based on eclipse width and the radial velocities of the irradiated hemeisphere on the donor.

\begin{table}
    \centering
    \begin{tabular}{l|c}
        Parameter & Value \\ \hline 
        \hline
        $P_{orb}$, min & $122.856\pm0.006$ \\ \hline
        $\Delta t_{ecl}$, s & $448.6\pm1.1$ \\ \hline
        $B_z$, MG & $15.1\pm 1.3$ \\ \hline
        $B_{cyc}$, MG & $\lesssim 45$ \\ \hline
        $T_{sh}$, keV & 8-18 \\ \hline
        $\beta$, $^\circ$ & 11 \\ \hline
        $\phi$, $^\circ$ & 31 \\ \hline
        $i$, $^\circ$ & $78.9\pm1.7$  \\ \hline
        $q$ & $0.33\pm 0.10$  \\ \hline
        $M_1$, $M_\odot$ & $0.57\pm 0.15$  \\ \hline
        $R_1$, $R_\odot$ & $0.013\pm0.002$  \\ \hline
        $M_2$, $M_\odot$ & $0.174\pm 0.022$  \\ \hline
        $R_2$, $R_\odot$ & $0.206\pm0.009$  \\ \hline
    \end{tabular}
    \caption{The main system parameters of the polar {\obj}}
    \label{tab:all_pars}
\end{table}

\section{Acknowledgements}
The observations with the SAO RAS telescopes are supported by the Ministry of Science and Higher Education of the Russian Federation. The instrumentation is updated within the “Science and Universities” National Project.

\clearpage


\end{document}